# Quantum entanglement in *n*-qubit real equally weighted states

Ri Qu, Yanru Bao

*School of Computer Science & Technology, Tianjin University, Tianjin 300072, China*

The *n*-qubit real equally weighted states are employed in some quantum algorithms including Deutsch-Jozsa, Grover, Simon, and so on. We qualitatively investigate the entanglement properties of *n*-qubit real equally weighted states. Firstly, all of the *n*-qubit real equally weighted states are classified into 7 parts by means of their structural degrees. Then we analyze the multipartite entanglement features of the states in every part by means of separable and similar degrees.

## I. INTRODUCTION

Entanglement is one of the most astonishing features of quantum mechanics. It is first recognized by Einstein, Podolsky and Rosen [1] and Schrödinger [2]. As known to us, quantum entanglement is a major resource in quantum communication, e.g. quantum dense coding [3], quantum teleportation [4], quantum key distribution [5], and so on. However it is not yet thoroughly clear whether quantum entanglement has an essential role for achieving computational speedup in current quantum algorithms. Recently, Bruß and Macchiavello [6] have elucidated the role of multipartite entanglement [7] in Deutsch-Jozsa [8], Grover [9], and Simon [10] algorithms by studying the properties of the so-called real equally weighted states (REWS's). They point out that multipartite entanglement is an important property in the above algorithms. However their analysis is limited for the special REWS's by some conditions, e.g., the REWS's generated by constant and balanced functions, etc. In this work, we qualitatively investigate the entanglement properties of all *n*-qubit REWS's and formally prove our results.

The paper is organized as follows. In Sec. II we review the definitions of REWS's and multipartite separability and entanglement in *n*-qubit pure states, and fix some notation. Moreover, we introduce several definitions including structural, separable, similar degrees, and so on. In Sec. III we first divide all of the *n*-qubit REWS's into 7 parts according to their structural degrees. Then we study the multipartite separability and entanglement properties of the states in every part by their separable and similar degrees. We summarize the results in Sec. IV.

## II. DEFINITIONS AND NOTATION

Let $\{|x\rangle \mid x=0,...,2^n-1\}$ be the computational basis of *n* qubits. Then we give the following definitions.

*Definition 1:* If an *n*-qubit state $|\psi_{REWS}^n\rangle$ is of the form

$$|\psi_{REWS}^n\rangle \equiv \frac{1}{\sqrt{2^n}} \sum_{x=0}^{2^n-1} a_x |x\rangle \qquad (1)$$

where $a_x \in \{1,-1\}$ for all *x*, then $|\psi_{REWS}^n\rangle$ is called an *n*-qubit *real equally weighted state*

(REWS).

*Definition 1':* If an $n$-qubit state $|\psi_{REWS}^n\rangle$ is of the form

$$|\psi_{REWS}^n\rangle \equiv \frac{1}{\sqrt{2^n}} \sum_{x=0}^{2^n-1} (-1)^{f(x)} |x\rangle \qquad (2)$$

where $f:\{0,...,2^n-1\} \to \{0,1\}$ is a Boolean function, then $|\psi_{REWS}^n\rangle$ is called an $n$-qubit REWS, and $f(x)$ is called the *relative function* of $|\psi_{REWS}^n\rangle$, i.e., $|\psi_{REWS}^n\rangle$ is generated by $f(x)$.

It is easily seen that both of the above two definitions are equivalent. Definition 1' is derived from the fact [6] that every REWS can be achieved by applying the unitary transformation $U_f$ corresponding to its relative function $f(x)$ to the special REWS $|\psi_+^n\rangle$ in the following form (5). For convenience, we denote an $n$-qubit REWS and the set of $n$-qubit REWS's by $|\psi_{REWS}^n\rangle$ and $W^n$, respectively. Notice that $|\psi_{REWS}^n\rangle$ and $e^{i\theta}|\psi_{REWS}^n\rangle$ are regarded as the same state in quantum theory. In this paper, there is only one exception that $|\psi_{REWS}^n\rangle$ and $-|\psi_{REWS}^n\rangle$ are thought as two different states. It is the reason that $|\psi_{REWS}^n\rangle$ and $-|\psi_{REWS}^n\rangle$ are different in some quantum algorithms. For example, in Grover algorithm [9], $|\psi_+^n\rangle$ means that the search problem has no solution, while $|\psi_-^n\rangle = -|\psi_+^n\rangle$ shows that the search problem has $2^n$ solutions. Thus the cardinal number of $W^n$ is equal to $2^{2^n}$.

*Definition 2:* Define by $\Delta(|\psi_{REWS}^n\rangle)$ the number of $a_x = -1$ inside (1) ($f(x) = 1$ in (2)), and $\Delta(|\psi_{REWS}^n\rangle)$ is called by the *structural degree* of $|\psi_{REWS}^n\rangle$.

Obviously, for all $|\psi_{REWS}^n\rangle$ we can obtain

$$\Delta(|\psi_{REWS}^n\rangle) \in \{0,1,...,2^n\}, \qquad (3)$$

$$\Delta(-|\psi_{REWS}^n\rangle) = 2^n - \Delta(|\psi_{REWS}^n\rangle). \qquad (4)$$

We can define some special REWS's by Definition 2 as follows. If $\Delta(|\psi_{REWS}^n\rangle)$ is odd (even), $|\psi_{REWS}^n\rangle$ is called by an odd (even) REWS. If $\Delta(|\psi_{REWS}^n\rangle) = 2^{n-1}$, $|\psi_{REWS}^n\rangle$ is called balanced. If $\Delta(|\psi_{REWS}^n\rangle) \in \{0, 2^n\}$, i.e., $|\psi_{REWS}^n\rangle$ is of the form

$$|\psi_\pm^n\rangle \equiv \pm\frac{1}{\sqrt{2^n}}\sum_{x=0}^{2^n-1}|x\rangle, \qquad (5)$$

then $|\psi_{REWS}^n\rangle$ is constant. Obviously, the REWS is balanced (constant) if and only if its relative function is balanced (constant). Further classification of even REWS's according to their structural degrees is shown in Sec. III.

The definition of separability and entanglement in multiqubit case is much richer than in biqubit case [7]. Suppose $|\psi^n\rangle$ is an *n*-qubit pure state. We define multipartite separability and entanglement of $|\psi^n\rangle$ as follows.

*Definition 3:* If $|\psi^n\rangle$ can be written as a tensor product of pure states of *k* individual subsystems, then $|\psi^n\rangle$ is called *k-separable*.

Denote by $S_k$ the set of pure *k*-separable states, then it is obvious that $S_n \subset ... \subset S_2 \subset S_1$.

*Definition 4:* If $|\psi^n\rangle \in S_k - S_{k+1}$ where $k \in \{1,2,...,n-1\}$, then $\delta(|\psi^n\rangle) \equiv k$ is called by the *separable degree* of $|\psi^n\rangle$. Moreover, $\delta(|\psi^n\rangle) = n$ if and only if $|\psi^n\rangle \in S_n$.

We can find the following properties about the separable degree by the above definition.

$$\delta(|\psi^n\rangle) = \delta(-|\psi^n\rangle), \qquad (6)$$

$$\delta(|\psi_1\rangle \otimes |\psi_2\rangle \otimes ... \otimes |\psi_k\rangle) = \sum_{i=1}^{k} \delta(|\psi_i\rangle) \qquad (7)$$

where $|\psi_1\rangle, |\psi_2\rangle, ..., |\psi_k\rangle$ are of pure states.

*Definition 5:* If $\delta(|\psi^n\rangle) = 1$, then $|\psi^n\rangle$ is called *fully entangled*. If $\delta(|\psi^n\rangle) = n$, then $|\psi^n\rangle$ is *fully separable*. If $n \geq 3$ and $\delta(|\psi^n\rangle) \in \{2,3,...,n-1\}$, then $|\psi^n\rangle$ is called by a *partially separable state*.

Let us consider the following case. Suppose $|\psi_1\rangle \equiv |\Phi^+\rangle \otimes |0\rangle$ and $|\psi_2\rangle \equiv |1\rangle \otimes |\Phi^+\rangle$ are two *3*-qubit states where the singlet state $|\Phi^+\rangle \equiv \frac{1}{\sqrt{2}}(|0\rangle|1\rangle + |1\rangle|0\rangle)$. It is known that the tensor product forms of $|\psi_1\rangle$ and $|\psi_2\rangle$ are similar since they are of the form $|\Phi^+\rangle \otimes |x\rangle$ where $x \in \{0,1\}$. Thus we give the following definitions which are used to assess the similarity between

two states.

*Definition 6:* Suppose $|\psi^n\rangle$ and $|\phi^n\rangle$ are two different *n*-qubit states. If there exists $k \in \{1,2,...,n-1\}$ such that $|\psi^n\rangle = |\varphi^k\rangle \otimes |\psi^{n-k}\rangle$, $|\phi^n\rangle = |\varphi^k\rangle \otimes |\phi^{n-k}\rangle$, then $|\psi^n\rangle$ is called *k-identical* with $|\phi^n\rangle$, and $|\varphi^k\rangle$ is called by their *k-identical part*. Otherwise $|\psi^n\rangle$ is *0-identical* with $|\phi^n\rangle$. Notice that $|\psi^n\rangle$ is *n-identical* with $e^{i\theta}|\psi^n\rangle$.

*Definition 7:* Suppose $|\psi^n\rangle$ and $|\phi^n\rangle$ are two *n*-qubit states. Define

$$\Gamma(|\psi^n\rangle,|\phi^n\rangle) \equiv \max_{k \in \{0,1,...,n\}} |\psi^n\rangle \text{ is } k\text{-identical with } |\phi^n\rangle. \qquad (8)$$

And $\Gamma(|\psi^n\rangle,|\phi^n\rangle)$ is called by the similar degree of $|\psi^n\rangle$ and $|\phi^n\rangle$.

According to Definition 6 and 7, we can find the following properties:

$$\Gamma(|\psi^n\rangle,|\phi^n\rangle) = \Gamma(|\phi^n\rangle,|\psi^n\rangle), \qquad (9)$$

$$\Gamma(|\psi^n\rangle,|\psi^n\rangle) = n, \qquad (10)$$

Finally we denote the set of integers by $Z$. Denote by $Z^+$ the set of positive integers.

## III. MULTIPARTITE SEPARABILITY AND ENTANGLEMENT

In order to analyze the separability and entanglement features of all *n*-qubit REWS's, we divide the states in $W^n$ into 7 parts: odd ones, constant ones, balanced ones, even ones with their structural degrees $\in [2, 2^{n/2})$, even ones with their structural degrees $\in (2^n - 2^{n/2}, 2^n - 2]$, even ones with their structural degrees $\in [2^{n/2}, 2^{n-1} - 2]$, and even ones with their structural degrees $\in [2^{n-1} + 2, 2^n - 2^{n/2}]$. The entanglement properties of the states in the first four parts have been indicated in [6]. But [6] are formally not proved some results while we prove them in this paper. Moreover, we also investigate the entanglement properties of the states in the last three parts which are not considered in [6]. Notice that some of the above 7 parts possibly include nothing for small *n*. For example, for $n = 2$ no REWS is of the last four parts, and for $n = 3$ no REWS is of the last two parts.

Now let us study the separability and entanglement properties of ones in every part. We first introduce the following lemma.

*Lemma 1:* Suppose $|\psi^n_{REWS}\rangle$ is *k*-separable, i.e.,

$$|\psi^n_{REWS}\rangle = |\psi^{l_1}\rangle \otimes |\psi^{l_2}\rangle \otimes ... \otimes |\psi^{l_k}\rangle \qquad (11)$$

where $\sum_{i=1}^{k} l_i = n$ and $k \in \{1, 2, ..., n\}$. Then

(i) For all $i \in \{1, 2, ..., k\}$, $|\psi^{l_i}\rangle$ is the $l_i$-qubit REWS, i.e., $|\psi_{l_i}\rangle \in W^{l_i}$.

(ii) $|\psi_{REWS}^n\rangle$ is balanced if and only if at least one $|\psi_{l_i}\rangle$ is balanced.

(iii) $|\psi_{REWS}^n\rangle$ is fully separable if and only if its relative function is one of

$$f(x) = ax \tag{12}$$

and

$$f(x) = ax \oplus 1 \tag{13}$$

where $ax \equiv a_1 x_1 \oplus a_2 x_2 \oplus \cdots \oplus a_n x_n$, and $a$ is an $n$-bit string.

Proof: We first prove (i). Suppose that

$$|\psi^{l_i}\rangle = \sum_{j=0}^{2^{l_i}-1} \alpha_j^{l_i} |j\rangle \text{ with } \sum_{j=0}^{2^{l_i}-1} \|\alpha_j^{l_i}\|^2 = 1. \tag{14}$$

Substituting (14) into (11), we obtain

$$|\psi_{REWS}^n\rangle = \sum_{j_1=0}^{2^{l_1}-1} \alpha_{j_1}^{l_1} |j_1\rangle \otimes \sum_{j_2=0}^{2^{l_2}-1} \alpha_{j_2}^{l_2} |j_2\rangle \otimes ... \otimes \sum_{j_k=0}^{2^{l_k}-1} \alpha_{j_k}^{l_k} |j_k\rangle. \tag{15}$$

By comparing (1) with (15), we can obtain that for all $j_1, j_2, ..., j_k$ the absolute value of $\alpha_{j_1}^{l_1} \cdot \alpha_{j_2}^{l_2} \cdot ... \cdot \alpha_{j_k}^{l_k}$ equals to $\frac{1}{\sqrt{2^n}}$, that is, $\left|\alpha_{j_1}^{l_1} \cdot \alpha_{j_2}^{l_2} \cdot ... \cdot \alpha_{j_k}^{l_k}\right| = \frac{1}{\sqrt{2^n}}$. Thus for all $|\psi^{l_i}\rangle$ we immediately get $|\alpha_0^{l_i}| = |\alpha_1^{l_i}| = ... = |\alpha_{2^{l_i}}^{l_i}| = \frac{1}{\sqrt{2^n}}$ apart from one global phase factor, which means $|\psi^{l_i}\rangle \in W^{l_i}$.

The proposition (ii) has been proved in [6]. Now let us prove (iii) as follows. (If) The state generated by (12) is fully separable since $\frac{1}{\sqrt{2^n}} \sum_{x=0}^{2^n-1} (-1)^{ax} |x\rangle = \bigotimes_{j=0}^{n} \frac{1}{\sqrt{2}} \left(|0\rangle + (-1)^{a_j} |1\rangle\right)$. For a fixed $a$, the state generated by (13) is the same as one generated by (12) apart from global phase factor -1. Thus the sate generated by (13) is also fully separable. (Only if) According to (i), $|\psi_{REWS}^n\rangle$ has to be written as one of the forms $\bigotimes_{j=0}^{n} \frac{1}{\sqrt{2}} \left(|0\rangle + (-1)^{a_j} |1\rangle\right)$ and $-\bigotimes_{j=0}^{n} \frac{1}{\sqrt{2}} \left(|0\rangle + (-1)^{a_j} |1\rangle\right)$ whose relative functions are respectively (12) and (13). □

Notice that for $a = 0$ the functions (12) and (13) are constant, and thus two REWS's corresponding to them are constant and fully separable. For $a \neq 0$, the functions (12) and (13) are balanced, and therefore the REWS's generated by them are balanced and fully separable. Thus any fully separable REWS must be either constant or balanced.

### A. Odd REWS's

In this section, we study the multipartite entanglement features of the odd REWS's.

*Theorem 2:*

(i) If $\Delta(|\psi_{REWS}^n\rangle)$ is odd, then $\delta(|\psi_{REWS}^n\rangle) = 1$, i.e., $|\psi_{REWS}^n\rangle$ is fully entangled.

(ii) Suppose $|\psi_{REWS}^n\rangle$ and $|\phi_{REWS}^n\rangle$ are odd. $|\psi_{REWS}^n\rangle$ and $|\phi_{REWS}^n\rangle$ is different if an only if $\Gamma(|\psi_{REWS}^n\rangle, |\phi_{REWS}^n\rangle) = 0$.

Proof: Assume that $|\psi_{REWS}^n\rangle$ would not be fully entangled. Then there would exist $k \in \{1, 2, ..., n-1\}$ such that $|\psi_{REWS}^n\rangle = |\psi_{REWS}^k\rangle \otimes |\psi_{REWS}^{n-k}\rangle$. By Definition 2,

$$\Delta(|\psi_{REWS}^n\rangle) = \left[2^k - \Delta(|\psi_{REWS}^k\rangle)\right] \cdot \Delta(|\psi_{REWS}^{n-k}\rangle) + \left[2^{n-k} - \Delta(|\psi_{REWS}^{n-k}\rangle)\right] \cdot \Delta(|\psi_{REWS}^k\rangle), \quad (16)$$

which would be even, while we require that $\Delta(|\psi_{REWS}^n\rangle)$ is odd. Thus $\delta(|\psi_{REWS}^n\rangle) = 1$. Form (i) we can easily obtain (ii). □

### B. Constant REWS's

In this section, we study the multipartite entanglement properties of $|\psi_{\pm}^n\rangle$ in (5). According to the above lemma and Definition 7, we can immediately get the following corollary.

*Corollary 3:*

(i) $\delta(|\psi_{\pm}^n\rangle) = n$, i.e., $|\psi_{\pm}^n\rangle$ are fully separable.

(ii) The similar degree of $|\psi_+^n\rangle$ and $|\psi_-^n\rangle$ is equal to *n-1*, i.e., $\Gamma(|\psi_+^n\rangle, |\psi_-^n\rangle) = n-1$. And their *n-1*-identical part can be written as the form $|\psi_+^{n-1}\rangle$.

### C. Balanced REWS's

In this section, we study the multipartite entanglement properties of balanced REWS's.

*Corollary 4:*

(i) If $\Delta(|\psi_{REWS}^2\rangle) = 2$, then $\delta(|\psi_{REWS}^2\rangle) = 2$, ie., $|\psi_{REWS}^2\rangle$ is fully separable.

(ii) Suppose $|\psi_{REWS}^2\rangle$ and $|\phi_{REWS}^2\rangle$ are balanced. $|\psi_{REWS}^2\rangle$ and $|\phi_{REWS}^2\rangle$ are different if and only if $\Gamma(|\psi_{REWS}^2\rangle, |\phi_{REWS}^2\rangle) = 1$ and their *1-identical* part can be written as the form $\frac{1}{\sqrt{2}}(|0\rangle - |1\rangle)$.

(iii) If $n \geq 3$, there exists $k \in \{1, 2, ..., n\}$ such that $\delta(|\psi_{REWS}^n\rangle) = k$.

Proof: Let us prove (i) and (iii). According to Lemma 1 (iii), the number of fully separable $n$-qubit REWS's is equal to $2 \cdot 2^n = 2^{n+1}$. Therefore, the number of fully separable and balanced $n$-qubit ones is $2^{n+1} - 2$ since there are two constant states that are fully separable. However the number of balanced REWS's is given by $B(2^n, 2^{n-1})$, where $B$ denotes the binomial coefficient. If $n = 2$, then $B(2^2, 2^{2-1}) = 2^{2+1} - 2 = 6$, and thus all of the balanced REWS's are fully separable. However, For $n \geq 3$ there are some balanced $n$-qubit ones that are not fully separable since $B(2^n, 2^{n-1}) > 2^{n+1} - 2$. The paper [6] has proved that the number of 2-separable balanced $n$-qubit ones is given by

$$\frac{1}{2}\sum_{k=1}^{n-1} B(n,k) \cdot B(2^k, 2^{k-1}) \left[ 2^{2^{n-k}} - \frac{1}{2} B(2^{n-k}, 2^{n-k-1}) \right]. \quad (17)$$

As $2^{n+1} - 2 < (17) < B(2^n, 2^{n-1})$, the set of balanced $n$-qubit ones will include fully entangled states. Moreover, the above set also includes partially separable states according to Lemma 1(ii). We can easily prove (ii) according to (i) and Lemma 1(ii). □

If two balanced REWS's $|\psi_{REWS}^n\rangle$ and $|\phi_{REWS}^n\rangle$ are partially separable and have the same separable degree, it is possible that $|\psi_{REWS}^n\rangle$ is 0-identical with $|\phi_{REWS}^n\rangle$. For example, for $n \geq 4$ we have the following states: $|\psi_{REWS}^n\rangle = \frac{1}{\sqrt{2}}(|0\rangle + |1\rangle) \otimes |\psi_{REWS}^{n-1}\rangle$ where $\delta(|\psi_{REWS}^{n-1}\rangle) = 1$ and $\Delta(|\psi_{REWS}^{n-1}\rangle) = 2^{n-2}$, and $|\phi_{REWS}^n\rangle = \frac{1}{\sqrt{2}}(|0\rangle - |1\rangle) \otimes |\phi_{REWS}^{n-1}\rangle$ where $\delta(|\phi_{REWS}^{n-1}\rangle) = 1$ and $\Delta(|\phi_{REWS}^{n-1}\rangle) = 1$. According to Lemma 1(ii), the above two states are balanced. Thus the state $|\psi_{REWS}^n\rangle$ is 0-identical with $|\phi_{REWS}^n\rangle$ though both of their separable degrees equal to 2.

### D. Even REWS's with their structural degrees $\in [2, 2^{n/2})$

The paper [6] has qualitatively analyzed the entanglement properties for the even REWS's with their structural degree $\in [2, 2^{n/2})$. In this section, we formally prove the conclusions in [6] and give some new results about separability and similarity of the above REWS's.

*Theorem 5:* Suppose $M = 2^q(2p+1) \in [2, 2^{n/2})$ where $p \in Z, q \in Z^+$. Suppose $|\psi_{REWS}^n\rangle$ is a

REWS whose structural degree is equal to $M$, that is, $\Delta(|\psi_{REWS}^n\rangle) = M$. Then

(i) For $k \in \{1, 2, ..., q\}$, $|\psi_{REWS}^n\rangle$ is $k+1$-separable if and only if $|\psi_{REWS}^n\rangle$ can be written as the form

$$|\psi_{REWS}^n\rangle = |\psi_{REWS}^k\rangle \otimes |\psi_{REWS}^{n-k}\rangle \tag{18}$$

where $\Delta(|\psi_{REWS}^k\rangle) = 0$ and $\Delta(|\psi_{REWS}^{n-k}\rangle) = 2^{q-k}(2p+1)$.

(ii) For $k \in \{1, 2, ..., q\}$, $\delta(|\psi_{REWS}^n\rangle) = k+1$ if and only if $|\psi_{REWS}^n\rangle$ can be written as the form

$$|\psi_{REWS}^n\rangle = |\psi_{REWS}^k\rangle \otimes |\psi_{REWS}^{n-k}\rangle \tag{19}$$

where $\Delta(|\psi_{REWS}^k\rangle) = 0$, $\Delta(|\psi_{REWS}^{n-k}\rangle) = 2^{q-k}(2p+1)$ and $\delta(|\psi_{REWS}^{n-k}\rangle) = 1$.

(iii) For any $k \in \{0, 1, ..., q\}$ there exists the REWS $|\phi_{REWS}^n\rangle$ such that $\Delta(|\phi_{REWS}^n\rangle) = M$ and $\delta(|\phi_{REWS}^n\rangle) = k+1$.

Proof: We firstly prove (i). (If) It is obvious. (Only if) As $|\psi_{REWS}^n\rangle$ is $k+1$-separable for $k \geq 1$, there exists $l \in \{1, 2, ..., n-1\}$ such that

$$|\psi_{REWS}^n\rangle = |\psi_{REWS}^l\rangle \otimes |\psi_{REWS}^{n-l}\rangle. \tag{20}$$

Thus the structural degree of $|\psi_{REWS}^n\rangle$ is given by

$$\left[2^{n-l} - \Delta(|\psi_{REWS}^{n-l}\rangle)\right] \cdot \Delta(|\psi_{REWS}^l\rangle) + \left[2^l - \Delta(|\psi_{REWS}^l\rangle)\right] \cdot \Delta(|\psi_{REWS}^{n-l}\rangle), \tag{21}$$

which follows that

$$\Delta(|\psi_{REWS}^l\rangle) \cdot \left[2^{n-l} - 2 \cdot \Delta(|\psi_{REWS}^{n-l}\rangle)\right] = 2^q(2p+1) - 2^l \cdot \Delta(|\psi_{REWS}^{n-l}\rangle). \tag{22}$$

Now Let us prove $2^{n-l} - 2 \cdot \Delta(|\psi_{REWS}^{n-l}\rangle) \neq 0$. Assume $2^{n-l} - 2 \cdot \Delta(|\psi_{REWS}^{n-l}\rangle) = 0$, then $|\psi_{REWS}^{n-l}\rangle$ would be balanced. According to Lemma 1(ii), $|\psi_{REWS}^n\rangle$ would also be balanced, which is contradictory with $\Delta(|\psi_{REWS}^n\rangle) \in [2, 2^{n/2})$. By (22), we obtain

$$\Delta(|\psi_{REWS}^l\rangle) = \frac{2^q(2p+1) - 2^l \cdot \Delta(|\psi_{REWS}^{n-l}\rangle)}{2^{n-l} - 2 \cdot \Delta(|\psi_{REWS}^{n-l}\rangle)}. \tag{23}$$

We analyze the tensor product form of $|\psi_{REWS}^n\rangle$ according to three situations as follows: (a) If

$2^q(2p+1) = 2^l \Delta\left(\left|\psi_{REWS}^{n-l}\right\rangle\right)$ then we have $\Delta\left(\left|\psi_{REWS}^l\right\rangle\right) = 0$ and $\Delta\left(\left|\psi_{REWS}^{n-l}\right\rangle\right) = 2^{q-l}(2p+1)$ where $l \in \{1, 2, ..., q\}$. Thus (20) is of the desired form (18).

(b) If $2^q(2p+1) > 2^l \Delta\left(\left|\psi_{REWS}^{n-l}\right\rangle\right)$ then $\Delta\left(\left|\psi_{REWS}^l\right\rangle\right) \geq 1$, which combined with (23) gives

$$(2-2^l)\Delta\left(\left|\psi_{REWS}^{n-l}\right\rangle\right) \geq 2^{n-l} - 2^q(2p+1). \tag{24}$$

Now let us prove that $\Delta\left(\left|\psi_{REWS}^{n-l}\right\rangle\right) = 0$. Assume that $\Delta\left(\left|\psi_{REWS}^{n-l}\right\rangle\right) \geq 1$, then it would imply that

$$2^{n-l} > 2^{\frac{n}{2}} > 2^q(2p+1) \tag{25}$$

since $1 \leq \Delta\left(\left|\psi_{REWS}^{n-l}\right\rangle\right) < 2^{q-l}(2p+1) < 2^{\frac{n}{2}-l}$. As $l \geq 1$, $(2-2^l)\Delta\left(\left|\psi_{REWS}^{n-l}\right\rangle\right) \leq 0$ which combined with (25) is contradictory with (24). So $\Delta\left(\left|\psi_{REWS}^{n-l}\right\rangle\right) = 0$. Substituting $\Delta\left(\left|\psi_{REWS}^{n-l}\right\rangle\right) = 0$ into (23), we obtain $\Delta\left(\left|\psi_{REWS}^l\right\rangle\right) = 2^{q+l-n}(2p+1)$ where $l \in \{n-q, n-q+1, ..., n-1\}$. Thus (20) is written as the form (18) as long as we let $k = n-l$.

(c) If $2^q(2p+1) < 2^l \Delta\left(\left|\psi_{REWS}^{n-l}\right\rangle\right)$, this can be transferred into situations (a) and (b) by means of the properties of structural degree. Substituting (23) into $\Delta\left(\left|\psi_{REWS}^l\right\rangle\right) \leq 2^l$, we obtain

$$2^{n-l} - 2^{q-l}(2p+1) \leq \Delta\left(\left|\psi_{REWS}^{n-l}\right\rangle\right) \leq 2^{n-l}. \tag{26}$$

According to (4), we can get $2^q(2p+1) \geq 2^l \cdot \Delta\left(-\left|\psi_{REWS}^{n-l}\right\rangle\right)$ that is the same with situations (a) and (b) for $-\left|\psi_{REWS}^{n-l}\right\rangle$ while (20) should be modified to

$$\left|\psi_{REWS}^n\right\rangle = \left(-\left|\psi_{REWS}^l\right\rangle\right) \otimes \left(-\left|\psi_{REWS}^{n-l}\right\rangle\right). \tag{27}$$

Note that, in fact, $\left|\psi_{REWS}^l\right\rangle \otimes \left|\psi_{REWS}^{n-l}\right\rangle = \left(-\left|\psi_{REWS}^l\right\rangle\right) \otimes \left(-\left|\psi_{REWS}^{n-l}\right\rangle\right)$. Thus (20) is also of the desired form (18), which is the same as for the situations (a) and (b).

Now we prove (ii) in the theorem. (If) Since $\left|\psi_{REWS}^n\right\rangle = \left|\psi_{REWS}^k\right\rangle \otimes \left|\psi_{REWS}^{n-k}\right\rangle$, we can obtain $\delta\left(\left|\psi_{REWS}^n\right\rangle\right) = \delta\left(\left|\psi_{REWS}^k\right\rangle\right) + \delta\left(\left|\psi_{REWS}^{n-k}\right\rangle\right)$ according to (7). Thus $\delta\left(\left|\psi_{REWS}^n\right\rangle\right) = k+1$. (Only if) As $k \geq 1$ and $\delta\left(\left|\psi_{REWS}^n\right\rangle\right) = k+1$, $\left|\psi_{REWS}^n\right\rangle$ is $k+1$-separable. According to (i), we have $\left|\psi_{REWS}^n\right\rangle = \left|\psi_{REWS}^k\right\rangle \otimes \left|\psi_{REWS}^{n-k}\right\rangle$ where $\Delta\left(\left|\psi_{REWS}^k\right\rangle\right) = 0$ and $\Delta(\left|\psi_{REWS}^{n-k}\right\rangle) = 2^{q-k}(2p+1)$. Form

(7), we can get $\delta(|\psi_{REWS}^n\rangle) = \delta(|\psi_{REWS}^k\rangle) + \delta(|\psi_{REWS}^{n-k}\rangle) = k + \delta(|\psi_{REWS}^{n-k}\rangle)$. Thus we can have $\delta(|\psi_{REWS}^{n-k}\rangle) = 1$.

At last, let us prove (iii). For an $n$-qubit system, the number of 2-separable REWS's whose structural degrees equal to $M$ is $nB(2^{n-1}, M/2)$ according to (i), while the number of all REWS's whose structural degrees are equal to $M$ is $B(2^n, M)$. There are some fully entangled ones whose structural degrees are $M$ because $nB(2^{n-1}, M/2) < B(2^n, M)$. In particular, the paper [6] has proved that the REWS whose structural degree equals to $M$ is nearly fully entangled for $M \square 2^n$. Given $k \in \{1, ..., q\}$, we can get $n - k \geq 2$. Further there exists the fully entangled state $|\phi_{REWS}^{n-k}\rangle$ with $\Delta(|\psi_{REWS}^{n-k}\rangle) = 2^{q-k}(2p+1)$ according to Theorem 2 and the above analysis. Thus we can construct the state $|\phi_{REWS}^n\rangle = |\phi_{REWS}^k\rangle \otimes |\phi_{REWS}^{n-k}\rangle$ where $\Delta(|\phi_{REWS}^k\rangle) = 0$. It is obvious that $\Delta(|\phi_{REWS}^n\rangle) = M$ and $\delta(|\phi_{REWS}^n\rangle) = k+1$ according to (ii). □

According to Definition 7 and Theorem 5, we can get the following corollary.

*Corollary 6:* Suppose $|\psi_{REWS}^n\rangle$ and $|\phi_{REWS}^n\rangle$ are two different even REWS's whose structural degrees $\in [2, 2^{n/2})$. Then

$$\Gamma(|\psi_{REWS}^n\rangle, |\phi_{REWS}^n\rangle) = \min\{\delta(|\psi_{REWS}^n\rangle), \delta(|\phi_{REWS}^n\rangle)\} - 1. \qquad (28)$$

**E. Even REWS's with their structural degrees $\in (2^n - 2^{n/2}, 2^n - 2]$**

For any even REWS $|\psi_{REWS}^n\rangle$, it is clear that $\Delta(|\psi_{REWS}^n\rangle) \in (2^n - 2^{n/2}, 2^n - 2]$ if and only if $\Delta(-|\psi_{REWS}^n\rangle) \in [2, 2^{n/2})$ according to (4). So $-|\psi_{REWS}^n\rangle$ is of the states shown in Sec. III D. The entanglement propertis of $-|\psi_{REWS}^n\rangle$ can be prescribed by Theorem 5. Thus we can get the following theorem.

*Theorem 7:* Let $M = 2^n - 2^q(2p+1) \in (2^n - 2^{n/2}, 2^n - 2]$ where $p \in Z, q \in Z^+$. Let $|\psi_{REWS}^n\rangle$ be a REWS whose structural degree is equal to $M$, that is, $\Delta(|\psi_{REWS}^n\rangle) = M$. Then

(i) For $k \in \{1, 2, ..., q\}$, $|\psi_{REWS}^n\rangle$ is $k+1$-separable if and only if $|\psi_{REWS}^n\rangle$ is of the form

$$|\psi_{REWS}^n\rangle = |\psi_{REWS}^k\rangle \otimes |\psi_{REWS}^{n-k}\rangle \tag{29}$$

where $\Delta(|\psi_{REWS}^k\rangle) = 2^k$ and $\Delta(|\psi_{REWS}^{n-k}\rangle) = 2^{q-k}(2p+1)$.

(ii) For $k \in \{1, 2, ..., q\}$, $\delta(|\psi_{REWS}^n\rangle) = k+1$ if and only if $|\psi_{REWS}^n\rangle$ is of the form

$$|\psi_{REWS}^n\rangle = |\psi_{REWS}^k\rangle \otimes |\psi_{REWS}^{n-k}\rangle \tag{30}$$

where $\Delta(|\psi_{REWS}^k\rangle) = 2^k$, $\Delta(|\psi_{REWS}^{n-k}\rangle) = 2^{q-k}(2p+1)$ and $\delta(|\psi_{REWS}^{n-k}\rangle) = 1$.

(iii) For any $k \in \{0, 1, ..., q\}$ there exists the REWS $|\phi_{REWS}^n\rangle$ such that $\Delta(|\phi_{REWS}^n\rangle) = M$ and $\delta(|\phi_{REWS}^n\rangle) = k+1$.

(iv) For $2^n - M \ll 2^n$, all of the $n$-qubit REWS's whose structural degrees are equal to $M$ are nearly fully entangled.

Proof: (i)-(iii) can be proved by means of Theorem 5. Let us prove (iv). The number of 2-separable REWS's whose structural degrees equal to $M$ is $nB(2^{n-1}, (2^n - M)/2)$ according to (i), while the number of all REWS's whose structural degrees equal to $M$ is $B(2^n, M) = B(2^n, 2^n - M)$. Thus $\lim_{n \to \infty} nB(2^{n-1}, (2^n - M)/2) / B(2^n, 2^n - M) \to 0$ for $2^n - M \ll 2^n$. □

According to Definition 7, Theorem 5 and Theorem 7, we can generalize Corollary 6 as follows.

*Corollary 8:* Suppose $|\psi_{REWS}^n\rangle$ and $|\phi_{REWS}^n\rangle$ are two different even $n$-qubit REWS whose structural degrees $\in [2, 2^{n/2}) \cup (2^n - 2^{n/2}, 2^n - 2]$. Then

$$\Gamma(|\psi_{REWS}^n\rangle, |\phi_{REWS}^n\rangle) = \min\{\delta(|\psi_{REWS}^n\rangle), \delta(|\phi_{REWS}^n\rangle)\} - 1. \tag{31}$$

**F. Even REWS's with their structural degrees $\in [2^{n/2}, 2^{n-1} - 2]$**

In this section, we qualitatively analyzed the multipartite entanglement properties for the even REWS's whose structural degrees $\in [2^{n/2}, 2^{n-1} - 2]$.

*Theorem 9:* Suppose $M = 2^q \in [2^{n/2}, 2^{n-1} - 2]$ where $q \in Z^+$. Suppose $|\psi_{REWS}^n\rangle$ is a REWS whose structural degree is equal to $M$, that is, $\Delta(|\psi_{REWS}^n\rangle) = M$. Then

(i) For $k \in \{1, 2, ..., q\}$, $|\psi_{REWS}^n\rangle$ is $k+1$-separable if and only if $|\psi_{REWS}^n\rangle$ is of the form

$$|\psi_{REWS}^n\rangle = |\psi_{REWS}^k\rangle \otimes |\psi_{REWS}^{n-k}\rangle \tag{32}$$

where $\Delta(|\psi_{REWS}^k\rangle) = 0$ and $\Delta(|\psi_{REWS}^{n-k}\rangle) = 2^{q-k}$.

(ii) For $k \in \{1, 2, ..., q\}$, $\delta(|\psi_{REWS}^n\rangle) = k+1$ if and only if $|\psi_{REWS}^n\rangle$ is of the form

$$|\psi_{REWS}^n\rangle = |\psi_{REWS}^k\rangle \otimes |\psi_{REWS}^{n-k}\rangle \tag{33}$$

where $\Delta(|\psi_{REWS}^k\rangle) = 0$, $\Delta(|\psi_{REWS}^{n-k}\rangle) = 2^{q-k}$ and $\delta(|\psi_{REWS}^{n-k}\rangle) = 1$.

(iii) For any $k \in \{0, 1, ..., q\}$ there exists the REWS $|\phi_{REWS}^n\rangle$ such that $\Delta(|\phi_{REWS}^n\rangle) = M$ and $\delta(|\phi_{REWS}^n\rangle) = k+1$.

Proof: We firstly prove (i). (If) It is obvious. (Only if) As $|\psi_{REWS}^n\rangle$ is $k+1$-separable for $k \geq 1$, there exists $l \in \{1, 2, ..., n-1\}$ such that

$$|\psi_{REWS}^n\rangle = |\psi_{REWS}^l\rangle \otimes |\psi_{REWS}^{n-l}\rangle. \tag{34}$$

Thus the structural degree of $|\psi_{REWS}^n\rangle$ is given by

$$\left[2^{n-l} - \Delta(|\psi_{REWS}^{n-l}\rangle)\right] \cdot \Delta(|\psi_{REWS}^l\rangle) + \left[2^l - \Delta(|\psi_{REWS}^l\rangle)\right] \cdot \Delta(|\psi_{REWS}^{n-l}\rangle), \tag{35}$$

which follows that

$$\Delta(|\psi_{REWS}^l\rangle) \cdot \left[2^{n-l} - 2 \cdot \Delta(|\psi_{REWS}^{n-l}\rangle)\right] = 2^q - 2^l \cdot \Delta(|\psi_{REWS}^{n-l}\rangle). \tag{36}$$

W can prove $2^{n-l} - 2 \cdot \Delta(|\psi_{REWS}^{n-l}\rangle) \neq 0$, which is the same as for Theorem 5 (i). So we obtain

$$\Delta(|\psi_{REWS}^l\rangle) = \frac{2^q - 2^l \cdot \Delta(|\psi_{REWS}^{n-l}\rangle)}{2^{n-l} - 2 \cdot \Delta(|\psi_{REWS}^{n-l}\rangle)}. \tag{37}$$

We analyze the tensor product form of $|\psi_{REWS}^n\rangle$ according to three situations as follows: (a) If $2^q = 2^l \Delta(|\psi_{REWS}^{n-l}\rangle)$ then we have $\Delta(|\psi_{REWS}^l\rangle) = 0$ and $\Delta(|\psi_{REWS}^{n-l}\rangle) = 2^{q-l}$ where $l \in \{1, 2, ..., q\}$. Thus (34) is of the form (32).

(b) If $2^q > 2^l \Delta(|\psi_{REWS}^{n-l}\rangle)$, then it is easily seen that

$$0 \leq \Delta(|\psi_{REWS}^{n-l}\rangle) < 2^{q-l} < 2^{n-l-1}. \tag{38}$$

Now let us prove that $\Delta(|\psi_{REWS}^{n-l}\rangle) = 0$. Assume $1 \leq \Delta(|\psi_{REWS}^{n-l}\rangle) < 2^{q-l}$, then there would exist $s, t \in Z$ such that $\Delta(|\psi_{REWS}^{n-l}\rangle) = 2^s(2t+1)$. By (38), we would get $0 \leq s < q-l < n-l-1$,

which means that $2^{n-1-l-s}, 2^{q-l-s} \in Z$, $2^{n-1-l-s} \geq 2$ and $2^{q-l-s} \geq 2$. Substituting $\Delta(|\psi_{REWS}^{n-l}\rangle) = 2^s(2t+1)$ into (37), we would obtain

$$\Delta(|\psi_{REWS}^{l}\rangle) = \frac{2^q - 2^{l+s}(2t+1)}{2^{n-l} - 2^{s+1}(2t+1)} \tag{39}$$

$$= 2^{l-1} \cdot \frac{2^{q-l-s} - (2t+1)}{2^{n-1-l-s} - (2t+1)}. \tag{40}$$

According to (38), we would get

$$0 \leq 2^{q-l-s} - (2t+1) \leq 2^{n-1-l-s} - (2t+2). \tag{41}$$

Since $2^{n-1-l-s} \geq 2$, so $2^{n-1-l-s} - (2t+1)$ is odd. By (40), $2^{n-1-l-s} - (2t+1)$ would divide $2^{q-l-s} - (2t+1)$, which combined with (41) would give $2^{q-l-s} = (2t+1)$. It implies $2^q = 2^l \cdot 2^s(2t+1)$ while we require $2^q > 2^l \Delta(|\psi_{REWS}^{n-l}\rangle)$. Thus $\Delta(|\psi_{REWS}^{n-l}\rangle) = 0$. From (37), we obtain $\Delta(|\psi_{REWS}^{l}\rangle) = 2^{q+l-n}$ where $l \in \{n-q, n-q+1, ..., n-1\}$. Thus (34) is of the form (32) as long as we let $k = n-l$. (c) If $2^q < 2^l \Delta(|\psi_{REWS}^{n-l}\rangle)$, this situation can be transferred into situations (a) and (b) the same as for Theorem 5 (i).

The proof for (ii) and (iii) are the same as for Theorem 5 (ii) and (iii), respectively. □

*Theorem 10:* Suppose $M = 2^q(2p+1) \in [2^{n/2}, 2^{n-1} - 2]$ where $p, q \in Z^+$. Then for any $k \in \{1, 2, ..., q\}$ there exists the REWS $|\psi_{REWS}^{n}\rangle$ such that $\Delta(|\psi_{REWS}^{n}\rangle) = M$ and $|\psi_{REWS}^{n}\rangle$ can be written as the following form

$$|\psi_{REWS}^{n}\rangle = |\psi_{REWS}^{k}\rangle \otimes |\psi_{REWS}^{n-k}\rangle \tag{42}$$

where $\Delta(|\psi_{REWS}^{k}\rangle) = 0$ and $\Delta(|\psi_{REWS}^{n-k}\rangle) = 2^{q-k}(2p+1)$.

Proof: By (42), $\Delta(|\psi_{REWS}^{k}\rangle \otimes |\psi_{REWS}^{n-k}\rangle) = \Delta(-|\psi_{REWS}^{k}\rangle) \cdot \Delta(|\psi_{REWS}^{n-k}\rangle) = 2^q(2p+1)$ is given. □

*Theorem 11:* Suppose $k \in \{1, 2, ..., n-1\}$, $|\psi_{REWS}^{k}\rangle$ is a $k$-qubit REWS and $|\psi_{REWS}^{n-k}\rangle$ is an $n$-$k$-qubit one. Moreover $\Delta(|\psi_{REWS}^{k}\rangle) \neq 0, 2^{k-1}, 2^k$ and $\Delta(|\psi_{REWS}^{n-k}\rangle) \neq 0, 2^{n-k-1}, 2^{n-k}$. We can

construct the $n$-qubit REWS $|\psi_{REWS}^n\rangle = |\psi_{REWS}^k\rangle \otimes |\psi_{REWS}^{n-k}\rangle$. If $\Delta(|\psi_{REWS}^n\rangle) \in [0, 2^{n-1}]$, then $|\psi_{REWS}^n\rangle$ is even and $\Delta(|\psi_{REWS}^n\rangle) \in [2^{n/2}, 2^{n-1} - 2]$.

Proof: Since $\Delta(|\psi_{REWS}^k\rangle) \notin \{0, 2^{k-1}, 2^k\}$, $\Delta(|\psi_{REWS}^{n-k}\rangle) \notin \{0, 2^{n-k-1}, 2^{n-k}\}$ and $|\psi_{REWS}^n\rangle$ is 2-separable, $|\psi_{REWS}^n\rangle$ has to be even and $\Delta(|\psi_{REWS}^n\rangle) \in [2^{n/2}, 2^{n-1} - 2]$ according to Corollary 3, 4 and Theorem 2, 5, 9, 10. □

### G. Even REWS's with their structural degrees $\in [2^{n-1} + 2, 2^n - 2^{n/2}]$

It is obvious that $\Delta(|\psi_{REWS}^n\rangle) \in [2^{n-1} + 2, 2^n - 2^{n/2}]$ means $\Delta(-|\psi_{REWS}^n\rangle) \in [2^{n/2}, 2^{n-1} - 2]$, just as prescribed in Sec. III E. Then the entanglement properties of $-|\psi_{REWS}^n\rangle$ can be prescribed by Theorem 9, 10 and 11. Thus we can have the following theorems.

*Theorem 12:* Suppose $M = 2^n - 2^q \in [2^{n-1} + 2, 2^n - 2^{n/2}]$ where $q \in Z^+$. Suppose $|\psi_{REWS}^n\rangle$ is a REWS whose structural degree is equal to $M$, that is, $\Delta(|\psi_{REWS}^n\rangle) = M$. Then

(i) For $k \in \{1, 2, ..., q\}$, $|\psi_{REWS}^n\rangle$ is $k+1$-separable if and only if $|\psi_{REWS}^n\rangle$ is of the form

$$|\psi_{REWS}^n\rangle = |\psi_{REWS}^k\rangle \otimes |\psi_{REWS}^{n-k}\rangle \tag{32}$$

where $\Delta(|\psi_{REWS}^k\rangle) = 2^k$ and $\Delta(|\psi_{REWS}^{n-k}\rangle) = 2^{q-k}$.

(ii) For $k \in \{1, 2, ..., q\}$, $\delta(|\psi_{REWS}^n\rangle) = k+1$ if and only if $|\psi_{REWS}^n\rangle$ is of the form

$$|\psi_{REWS}^n\rangle = |\psi_{REWS}^k\rangle \otimes |\psi_{REWS}^{n-k}\rangle \tag{33}$$

where $\Delta(|\psi_{REWS}^k\rangle) = 2^k$, $\Delta(|\psi_{REWS}^{n-k}\rangle) = 2^{q-k}$ and $\delta(|\psi_{REWS}^{n-k}\rangle) = 1$.

(iii) For any $k \in \{0, 1, ..., q\}$ there exists the REWS $|\phi_{REWS}^n\rangle$ such that $\Delta(|\phi_{REWS}^n\rangle) = M$ and $\delta(|\phi_{REWS}^n\rangle) = k+1$.

*Theorem 13:* Suppose $M = 2^n - 2^q(2p+1) \in [2^{n-1} + 2, 2^n - 2^{n/2}]$ where $p, q \in Z^+$. Then for any $k \in \{1, 2, ..., q\}$ there exists the REWS $|\psi_{REWS}^n\rangle$ such that $\Delta(|\psi_{REWS}^n\rangle) = M$ and $|\psi_{REWS}^n\rangle$ can be written as the following form

$$|\psi_{REWS}^n\rangle = |\psi_{REWS}^k\rangle \otimes |\psi_{REWS}^{n-k}\rangle \tag{42}$$

where $\Delta(|\psi_{REWS}^{k}\rangle) = 2^{k}$ and $\Delta(|\psi_{REWS}^{n-k}\rangle) = 2^{q-k}(2p+1)$.

*Theorem 14:* Suppose $k \in \{1, 2, ..., n-1\}$, $|\psi_{REWS}^{k}\rangle$ is a $k$-qubit REWS and $|\psi_{REWS}^{n-k}\rangle$ is an $n$-$k$-qubit one. Moreover $\Delta(|\psi_{REWS}^{k}\rangle) \neq 0, 2^{k-1}, 2^{k}$ and $\Delta(|\psi_{REWS}^{n-k}\rangle) \neq 0, 2^{n-k-1}, 2^{n-k}$. We can construct the REWS $|\psi_{REWS}^{n}\rangle = |\psi_{REWS}^{k}\rangle \otimes |\psi_{REWS}^{n-k}\rangle$. If $\Delta(|\psi_{REWS}^{n}\rangle) \in [2^{n-1}+1, 2^{n}]$, then $|\psi_{REWS}^{n}\rangle$ is even and $\Delta(|\psi_{REWS}^{n}\rangle) \in [2^{n-1}+2, 2^{n} - 2^{n/2}]$.

## IV CONCLUSIONS

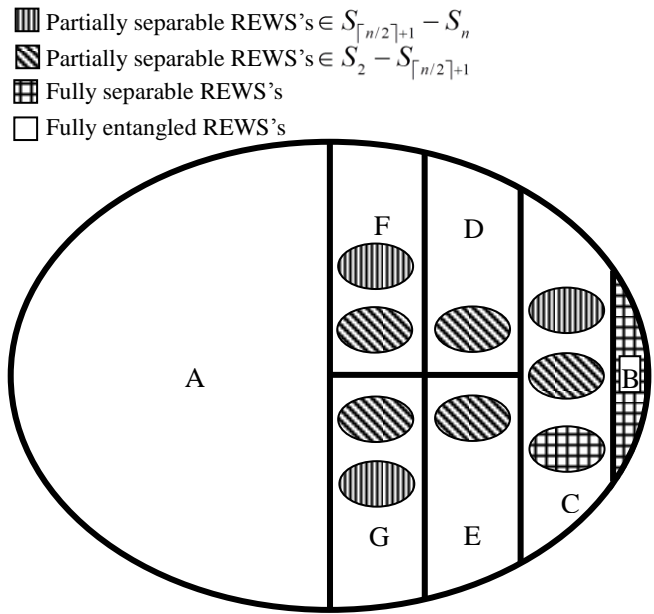

Fig. 1. Relation between their structural degrees and multipartite entanglement in $n$-qubit REWS's for $n \geq 4$. Respectively, A-G represent the sets which are formed by the states prescribed in Sec. III A-G.

In this paper, we qualitatively study the multipartite entanglement properties of all $n$-qubit REWS's. In particular, we investigate the entanglement properties of the states in the following parts: even REWS's with their structural degrees $\in (2^{n} - 2^{n/2}, 2^{n} - 2]$, even ones with their structural degree $\in [2^{n/2}, 2^{n-1} - 2]$, and even ones with their structural degrees $\in [2^{n-1} + 2, 2^{n} - 2^{n/2}]$, which are not considered in [6]. We get some results which are shown in Fig. 1. The set of $n$-qubit REWS's $W^{n}$ ($n \geq 4$) is partitioned into 7 subsets according to Sec. III. Every state in A is fully entangled while there are only two states in B and they are fully separable. C includes the fully separable, fully entangled and partially separable states in the certainty. Since D (F) is identical with E (G) apart from global phase factor -1, D (F) and E (G) have the same entanglement properties. D includes the fully entangled and partially separable states, but there is no fully separable one in D, which is the same as F. However $\max_{|\psi_{REWS}^n\rangle \in F} \delta(|\psi_{REWS}^n\rangle)$ is *n-1* while $\max_{|\psi_{REWS}^n\rangle \in D} \delta(|\psi_{REWS}^n\rangle)$ is $\lceil n/2 \rceil$, which implies that D does not includes $\lceil n/2 \rceil+1$-separable, $\lceil n/2 \rceil+2$-separable, …, and *n-1*-separable states but F does. For E and G, we can get analogous results. Moreover, we also discuss possible tensor product forms of the states in every part and study the similar degrees among the states by means of the forms.

In this work, we have not used any quantitative tools to analyze the multipartite entanglement properties of the REWS's, which will be the object of future investigation.

## ACKNOWLEDGMENTS

This work was financially supported by the National Natural Science Foundation of China under Grant No. 61170178.

**Reference**

[1] A. Einstein, B. Podolsky, and N. Rosen, Phys. Rev. 47, 777 (1935).
[2] E. Schrödinger, Naturwisssenschaften 23, 807 (1935).
[3] C. H. Bennett and S. J. Wiesner, Phys. Rev. Lett. 69, 2881 (1992).
[4] C. H. Bennett, G. Brassard, C. Crepeau, R. Jozsa, A. Peres, and W. K. Wooters, Phys. Rev. Lett. 70, 1895 (1993).
[5] C. H. Bennett, D. P. DiVincenzo, J. A. Smolin, and W. K. Wootters, Phys. Rev. A 54, 3824 (1996).
[6] D. Bruß and C. Macchiavello, Phys. Rev. A 83, 052313 (2011).
[7] R. Horodecki, P. Horodecki, M. Horodecki, and K. Horodecki, Rev. Mod. Phys. 81, 865 (2009).
[8] D. Deutsch and R. Jozsa, Proc. R. Soc. London A 439, 553 (1992).
[9] L. Grover, *Proceedings of the 28th Annual ACM Symposium on the Theory of Computing* (ACM Press, New York, 1996), p. 212.
[10] D. Simon, *in Proceedings of the 35th IEEE Symposium on the Foundations of Computer Science*, edited by S. Goldwasser (IEEE Computer Society, Los Alamitos, CA, 1994), p. 116.